\documentclass[11pt,letterpaper,twocolumn]{article}
\usepackage[margin=1in]{geometry}
\usepackage[T1]{fontenc} 

\usepackage{times}
\usepackage{helvet}
\usepackage{courier}

\usepackage[hyphens]{url}
\usepackage{microtype} 

\usepackage{graphicx}
\usepackage{natbib}
\usepackage{caption}

\usepackage{authblk}

\usepackage[hidelinks]{hyperref}

\usepackage{amsmath}
\usepackage{amssymb}
\usepackage[export]{adjustbox}
\usepackage[table]{xcolor}
\usepackage{longtable}
\usepackage{array} 

\usepackage{orcidlink}
\definecolor{orcidlogocol}{HTML}{A6CE39} 

\newcolumntype{L}[1]{>{\raggedright\arraybackslash}p{#1}} 
\newcolumntype{R}[1]{>{\raggedleft\arraybackslash}p{#1}}  

\definecolor{nslblue}{HTML}{0081B5}
\definecolor{nslgreen}{HTML}{4E946C}
\definecolor{nslorange}{HTML}{D05504}
\definecolor{nslred}{HTML}{91302C}
\title{Whose Polarization Is It? Auditing Fairness in Network Polarization Measures}

\author[1]{Stanis{\l}aw St\k{e}pie\'{n}\,\textcolor{orcidlogocol}{\orcidlink{0000-0003-2401-256X}}}
\author[1]{Michalina Janik\,\textcolor{orcidlogocol}{\orcidlink{0009-0009-6901-2644}}}
\author[2]{Akrati Saxena\,\textcolor{orcidlogocol}{\orcidlink{0000-0002-7151-6309}}}
\author[1]{Rados{\l}aw Michalski\,\textcolor{orcidlogocol}{\orcidlink{0000-0002-0106-655X}}}

\affil[1]{Wroclaw University of Science and Technology, Wroc{\l}aw, Poland}
\affil[2]{Leiden University, Leiden, The Netherlands}

\affil[ ]{\texttt{stanislaw.stepien@pwr.edu.pl}}

\begin{document}
\maketitle

\begin{abstract}
A polarization score is supposed to describe a community, not its surroundings. Yet when a minority community is embedded in a larger network, its measured polarization can rise or fall with the size and structure of the majority around it. We ask whether polarization measures are therefore fair to minorities and develop an audit built around two criteria: measurement invariance (MI), which asks whether otherwise equivalent majorities and minorities receive equivalent scores, and structural exogeneity (SE), which asks whether a minority's score changes when its surrounding majority changes. Resolution sensitivity (RS) is treated as a prerequisite for meaningful fairness assessment. We audit seven polarization measure families spanning the three network-operationalizable dimensions of polarization, introducing Network-Weighted Distance (NWD) to fill an unserved extremeness-separation combination. Across controlled synthetic networks, mechanism-isolating diagnostic topologies, and real political and communication networks, no measure passes all applicable checks. Minorities are systematically scored as more polarized than equivalent majorities. We trace this dependence to two separate channels: majority structure can alter the opinions that opinion-reading measures consume, and, on a different path, opinion-orthogonal structural change can affect the arithmetic of structural measures. Reading opinions is thus double-edged, letting contamination in when structure moves opinions but shielding the score when structure changes and opinions do not. We close with practical guidance for reporting polarization scores without inheriting these biases.
\end{abstract}

\section{Introduction}
\label{intro}
Consider two communities embedded in the same social network: a large political majority and a small, structurally peripheral minority. Suppose nothing about the minority changes - its internal opinions, its internal ties, and its connections to the majority all remain exactly as they were - while the majority simply grows, or its internal connections grow denser. Under several widely used polarization measures, the minority's reported score can shift even when the minority itself is unchanged, provided its opinions were shaped by the surrounding network rather than fixed in advance. As we show, this is the realistic case: when opinions form through social influence, majority structure leaks into the minority's score, whereas under externally fixed opinions the same measures are well behaved. The instrument meant to characterize the minority is, in part, measuring the majority.

Such scores are used to characterize echo chambers in online discourse \citep{Adamic2005}, track political division, and motivate platform-level and policy interventions. If the instrument is not exogenous to the groups it claims to measure, those claims inherit the contamination: a minority's apparent polarization becomes partly a property of the surrounding network.

Fairness has not been absent from network science, but it has so far been asked of a different target: whether \emph{interventions} treat groups equitably - homophily-reducing recommendations that disproportionately burden minority users \citep{Stoica2019}, or covering objectives that leave minority subgraphs under-served \citep{Rahmattalabi2019} - or whether \emph{predictive models} built on network and opinion data err more often for some groups than others \citep{Stepien2026}. However, we are not aware of work asking about the fairness of the measurement instruments themselves. The closest prior work, \cite{Salloum2022}, shows that several structural measures are sensitive to the relative sizes of the groups being compared, but treats this as a calibration problem rather than a symptom of a deeper structural issue (developed in the Literature Review). We intend to look at this problem through the lens of the measure's fairness and not as an issue of calibration.

In this paper we give fairness in polarization measurement an operational definition, resting on two criteria - measurement invariance (MI) and \emph{structural exogeneity} (SE) - with resolution sensitivity (RS) as a prerequisite that must hold before either can be assessed. To cover the one combination of dimensions that no established measure serves, extremeness together with separation, we introduce the Network-Weighted Distance (NWD). We then put the measures through a three-track audit: controlled parametric sweeps, mechanism-isolating diagnostic topologies, and corroboration on real-world networks. Its central finding is structural: the majority's context reaches a minority's score along two separate paths, by shaping the opinions an opinion-reading measure consumes and by feeding the structural computation of a structure-reading measure, and a measure's reliance on opinions is precisely what leaves it exposed on the first path and protects it on the second.

\section{Literature Review}\label{sec:relatedWorks}

\subsection{Polarization Measures and Their Dimensions}
\label{sec:polarizationMeasures}

Although we often speak of polarization as a single property, it is better understood as several distinct phenomena at once. Following \cite{BramsonPolarization}, we organize the measures around three network-operationalizable dimensions: \emph{extremeness} (D1), \emph{clustering} (D2), and \emph{separation} (D3). Extremeness captures how far opinions lie toward opposing ends of the opinion scale; clustering captures how strongly similar opinions or groups are organized in the network; and separation captures how strongly opposing groups are structurally divided. \cite{Hohmann2023} independently distinguish opinion, structural, and mesolevel-interplay components that map onto these dimensions. We use these dimensions to compare measures by what they capture, rather than treating polarization as a single quantity.

\begin{figure}[t]
\centering
\includegraphics[width=0.85\linewidth]{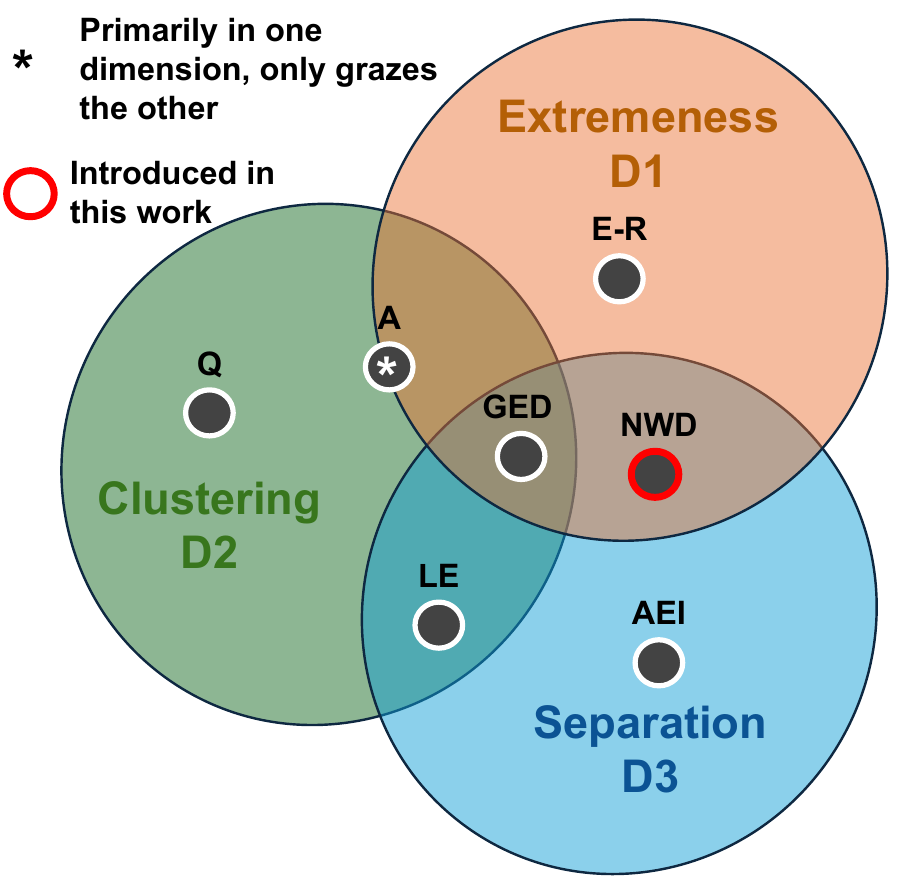}
\caption{Coverage of the seven audited measures across the three Bramson dimensions: extremeness (D1), clustering (D2), and separation (D3). Each measure sits in the region of the dimensions it captures. Modularity appears here as a single measure (Q) but is audited in two operationalisations, $Q_\text{opinion}$ and $Q_\text{struct}$.}
\label{fig:dim_coverage}
\end{figure}

We therefore audit measures spanning the dimensions and their combinations, mapped in Figure~\ref{fig:dim_coverage}. The distributional tradition is represented by the Esteban-Ray index (E-R) \citep{estebanRay1994}, which operates on the marginal opinion distribution and discards topology entirely (D1 only). The structural tradition centers on modularity (Q) \citep{Newman_2006}, whose single score hides a distinction that matters for our audit, so we split it into two operationalisations; with this split the seven measure families below yield eight evaluated scores. $Q_\text{opinion}$ scores the grouping defined by opinion sign: it reads opinions directly, and because that grouping is fixed rather than detected, it escapes modularity's resolution limit. $Q_\text{struct}$ instead ignores opinions, scoring the communities that the Leiden algorithm \citep{Traag2019} detects from structure alone at unit resolution, so it inherits the resolution limit \citep{Fortunato2007}.

Assortativity (A) \citep{Newman2003} measures opinion correlation across edges and is clustering-primary (D2). The Laplacian disagreement energy (LE) $\mathbf{x}^\top L\,\mathbf{x}$ (D2+D3) has the most direct opinion-dynamics grounding of the set but conflates conflict with diversity. The Adaptive E-I index (AEI) \citep{Chen2021AEI} targets separation alone through edge densities (D3), while the Generalized Euclidean Distance (GED) \citep{Hohmann2023} is the only established measure sensitive to all three dimensions at once (D1+D2+D3).

\citet{Guerra2013} likewise argue that community structure alone does not necessarily capture antagonism, motivating a polarization measure based explicitly on the boundaries between opposing communities. This measurement-oriented perspective complements ours: where they ask whether a measure captures the intended phenomenon, we ask whether it measures that phenomenon consistently when the surrounding network changes.

Because the three dimensions capture distinct properties, auditing only GED would obscure dimension-specific failures. Each dimension answers a different question about polarization, so the audit must span them rather than collapse them into one measure. This coverage analysis reveals one unserved established combination, D1+D3 (extremeness + separation), motivating the new measure introduced in the methodology.

Figure~\ref{fig:dim_coverage} maps the resulting coverage. It also shows the other unserved pairwise slot, D1+D2: the closest candidate, the dipole measure of \cite{Morales2015}, performs poorly after null-model normalization \citep{Salloum2022}. We therefore audit explicit coverage of the three dimensions rather than relying on any single “all-purpose” polarization score.

\subsection{Fairness in Polarization Measurement}\label{sec:fairnessBackground}
What happens if the \emph{measurement} of polarization is itself biased? Despite extensive work on fairness in network algorithms, we are not aware of prior work asking whether a polarization measure systematically assesses minority groups differently from otherwise equivalent majorities. On the polarization-\emph{intervention} problem, homophily-reducing recommendations \citep{Stoica2019} and unconstrained network objectives \citep{Rahmattalabi2019} are shown to disadvantage minorities systematically, and works like \citep{Stepien2026} find that certain opinion-prediction models misclassify structural and demographic minorities at higher rates. The emerging fairness-in-SNA literature surveyed by \cite{Saxena2024} spans link prediction, centrality, influence maximization, and community detection - all algorithmic outputs. Even fair community detection \citep{deVink2025}, the closest case, asks whether minority communities are recovered reliably, not whether resulting descriptive scores are equitable. Fairness of the \emph{measurement} itself, whether the instruments produce consistent assessments across structurally distinct subpopulations, is the gap we address.

The psychometric notion of measurement invariance \citep{VandenbergLance2000} is the closest analogue, and we adopt its construct-validity framing from \cite{JacobsWallach2021}. However, psychometric invariance was developed for comparing measurements across groups treated separately, whereas in a network the groups are coupled: a minority community's measured polarization can change when the surrounding majority changes, even if the minority itself does not. This \textbf{structural non-exogeneity} arises naturally when subgroup-level quantities depend on network-wide processes, and creates a fairness problem not addressed by either tradition. Related work by \cite{Gaitonde2020} shows that Friedkin-Johnsen equilibrium outcomes for a target group can shift when the surrounding topology is perturbed, while \cite{Salloum2022} documents a similar dependence on majority size. We formalize these effects as a fairness problem and evaluate seven polarization measures using two fairness criteria, structural exogeneity (SE) and measurement invariance (MI), together with a resolution-sensitivity (RS) prerequisite: a measure that cannot reliably detect small communities cannot meaningfully be tested for SE or MI.

Other fairness questions, including affective, temporal, and opinion-leader fairness, remain outside our scope.

We position our \emph{measurement audit} relative to the closest existing work on polarization-measure comparison, \cite{Salloum2022}. They show that eight structural polarization measures produce elevated scores on random networks with no planted group structure and note that several are sensitive to the relative sizes of the groups analyzed. We take this as our empirical starting point but depart in three respects: we read size-sensitivity as a symptom of structural non-exogeneity (a well-calibrated measure can still be non-exogenous); we ground the evaluation in opinion-dynamics theory, using analytically derived opinion assignments rather than static structure alone; and we add purpose-built diagnostic topologies that isolate the \emph{mechanism} of failure. Their analysis is also of whole networks, whereas ours is of communities embedded in larger ones, where surrounding structure can move a minority's score.

\section{Methodology}\label{sec:methodology}

\subsection{Overview and Rationale}
\label{sec:overviewandrationale}

We audit seven polarization measures, using three complementary experimental tracks. Track~A provides controlled parametric sweeps over network structure, group size, and opinion assignment; Track~B uses diagnostic topologies designed to isolate specific failure mechanisms; Track~C tests whether the synthetic patterns have plausible counterparts in real political and communication networks. The synthetic tracks carry the primary evidentiary weight because only they separate measurement effects from genuine changes in opinion and structure.

\subsection{Audited Measures}\label{sec:auditedMeasures}

The audit covers seven measure families, with modularity evaluated in two operationalisations, yielding eight evaluated scores. Together they span the three polarization dimensions and their combinations, including the new Network-Weighted Distance (NWD), introduced to fill the D1+D3 coverage gap.

NWD weights each pair's opinion distance by their $t$-step random-walk interaction probability:
\begin{equation}
    \text{NWD} = \frac{1}{n} \sum_i \sum_j P^t[i,j] \cdot |x_i-x_j|,
\end{equation}
where $P$ is the row-normalized transition matrix; we set $t=3$ throughout and leave sensitivity to this choice to future work. NWD captures extremeness through $|x_i-x_j|$ and separation through walk probabilities that decay across sparse inter-community bridges. Its per-agent decomposition $\delta_i(\text{NWD})=\sum_j P^t[i,j]|x_i-x_j|$ measures the experienced opinion distance that moderate brokers dilute in Topology~T7.

The resulting set deliberately contrasts measures that read opinions directly with measures that derive polarization from network structure.

\subsection{Fairness Evaluation Framework}\label{sec:farinessCriteria}
We define fairness operationally through two criteria - measurement invariance (MI) and structural exogeneity (SE) - together with a resolution-sensitivity (RS) prerequisite. MI tests whether equivalent groups receive equivalent scores, SE tests whether a group's score depends on its surrounding structure, and RS tests whether the group is detectable well enough for the other checks to be meaningful. The criteria are grounded in the psychometric tradition of measurement invariance \citep{VandenbergLance2000}; standard machine-learning fairness criteria do not transfer directly because descriptive measures lack ground-truth labels and error distributions, while network groups are structurally coupled. The framework is applied to all eight evaluated score operationalisations, although AEI is not substantively evaluable under the single-community minority design. The audit makes one interpretive assumption: we treat a change in a minority's score caused by changes in majority-induced opinions as a fairness problem. This is an assumption about comparability, not a claim that social influence is undesirable. Our reporting guidance does not depend on taking a position on that broader question.

\subsubsection{Measurement Invariance (MI)}\label{sec:measurementInvariance}
We operationalize \textbf{measurement invariance} by asking whether otherwise-equivalent majority and minority groups receive systematically different scores. The groups share the same opinion-generation scheme and structural probabilities, but differ in size and group role, so group size is treated as a nuisance attribute rather than a source of substantive polarization. We flag a violation only when their score distributions differ significantly (two-sample Kolmogorov--Smirnov test, Benjamini--Yekutieli corrected) \emph{and} the standardized gap exceeds $|d|=0.3$.
 A measure that systematically under- or over-reports for minorities gives an inequitable assessment regardless of how well its global score is calibrated. MI is evaluated in Track~A by comparing score distributions across majority and minority SBM blocks over the size ratio $r = n_\text{min}/n_\text{maj}$, and in Track~C by testing whether a minority community's contribution to the full-network score is proportional to its relative size and polarization intensity.

\subsubsection{Structural Exogeneity (SE)}\label{sec:structuralExogeneity}
\textbf{Structural exogeneity} is the central criterion of this audit. A measure $P$ satisfies it with respect to a fixed minority subgraph $G_\text{min}$ if the score is invariant to perturbations of the majority. Writing $\boldsymbol{\theta}_\text{maj} = (n_\text{maj},\, p_\text{in}^\text{maj},\, p_\text{out})$ for the majority's size, internal density, and inter-block connectivity, and holding the minority parameters $\boldsymbol{\theta}_\text{min}$ fixed,
\begin{equation}
    \frac{\partial}{\partial \boldsymbol{\theta}_\text{maj}}\,
    \mathbb{E}\bigl[P(G_\text{min}) \mid
    \boldsymbol{\theta}_\text{min},\boldsymbol{\theta}_\text{maj}\bigr]
    \approx \mathbf{0};
\end{equation}
that is, expected minority scores do not co-vary with majority structural parameters. A measure that fails assigns minority scores that depend partly on majority structure. MI does not imply SE: a measure can treat majority and minority groups symmetrically yet still depend on the surrounding majority.

We operationalize SE with a factorial sweep of the majority parameters in the SBM family, holding the minority fixed ($n_\text{min}=100$, $p_\text{in}^\text{min}=0.3$) and varying $n_\text{maj}$, $p_\text{in}^\text{maj}$, and $p_\text{out}$ over the grid given in the appendix, with 50 replicates per cell. In this experiment, each minority is scored on its own extracted subgraph, so majority structure can reach the minority score only through the opinions the minority holds. The grid is therefore run under two opinion-assignment schemes that isolate the two possible channels. In the \emph{bimodal} condition, opinions are drawn independently of the network and held fixed as majority structure varies; this is a control, since the majority is absent from the minority's scoring input. In the \emph{Friedkin--Johnsen} condition, the stubborn-opinion vector is fixed but equilibrium opinions are recomputed on each graph, allowing majority structure to affect the minority's opinions through social influence. Because the minority is still scored in isolation, any dependence in this condition must enter through opinion formation rather than through the measure's arithmetic.

For each measure, we regress minority scores on the three majority parameters and flag a structural-exogeneity violation when a predictor has a reliable, non-trivial effect. The Adaptive E-I Index is the exception: on these single-community minority subgraphs it returns a constant and is therefore \emph{untestable}, not exogenous.

\subsubsection{Resolution Sensitivity (RS)}
\textbf{Resolution sensitivity} is a prerequisite check rather than a fairness criterion proper: a measure that cannot detect polarization in small or peripheral communities at all fails through blindness, not inequity. A measure is resolution-sensitive with respect to $G_\text{min}$ if
\begin{equation}
    P(G_\text{min} \mid \text{polarized}) > P_\text{noise} + \delta,
\end{equation}
where $P_\text{noise}$ is the score assigned to the uniform baseline T6 - an Erdős–Rényi graph with uniformly distributed opinions - and $\delta$ is half a standard deviation of the null distribution. A measure that fails below a given size threshold cannot be assessed for the other two criteria at that scale, because any MI or SE clearance it then obtains is uninterpretable. The resolution limit of modularity established by \cite{Fortunato2007} gives a theoretical prediction for where $Q$ fails this check as a function of network size, which we verify empirically and use as a reference for the other measures.

\subsubsection{Measurement Invariance versus Resolution Sensitivity}
 Measurement invariance and resolution sensitivity can pull in opposite directions, since confident detection is harder for smaller groups. We separate \emph{uncertainty-driven} under-detection (high variance but an equivalent expected score, a finite-sample property and not a fairness failure) from \emph{bias-driven} under-detection (a systematically lower expected minority score after accounting for variance), flagging only the latter.

\subsubsection{Standard Network Topologies (Track A)}\label{sec:TrackA}
Track~A uses three network families. \emph{Erd\H{o}s--R\'enyi} graphs provide a structurally neutral baseline, with the T6 condition serving as the noise-floor reference for scores under uniformly distributed opinions. \emph{Barab\'asi--Albert} graphs add degree heterogeneity, whose hubs bridge opinion groups and dominate network-weighted measures. \emph{Stochastic block model} graphs with two blocks are the primary vehicle for all three criteria: fixing every structural parameter and varying only the size ratio $r = n_\text{min}/n_\text{maj} \in \{0.1, 0.2, 0.5, 1.0\}$ isolates the effect of group size. Parameter ranges are in the appendix.

Opinions in $[-1,1]$ are assigned by four schemes. \emph{Uniform} ($\mathcal{U}(-1,1)$) is the null assignment and fixes the T6 noise floor. \emph{Bimodal} draws from a two-Gaussian mixture at $\pm\mu$, and directly tests extremeness-sensitive measures like Esteban-Ray. \emph{Majority-skewed} gives the majority block bimodal opinions and the minority uniform, testing whether a measure attributes low polarization to the minority rather than absorbing it into the majority signal. \emph{FJ-equilibrium} computes opinions from the closed-form Friedkin-Johnsen equilibrium,
\begin{equation}
    \mathbf{x}^* = (I - \alpha \hat{A})^{-1}(1 - \alpha)\mathbf{s},
\end{equation}
with $\hat{A}$ the row-normalized adjacency, $\mathbf{s} \sim \mathcal{N}(\pm \mu_s, \sigma_s)$ a bimodal stubborn-opinion vector (by block in SBM, at random otherwise), and $\alpha$ the susceptibility to influence; its disagreement energy is exactly what the Laplacian energy minimizes \citep{musco2017minimizingpolarizationdisagreementsocial}. Parameter ranges are in the appendix, and each network-assignment pair is replicated 50 times.

\subsubsection{Diagnostic Topologies (Track B)}\label{sec:TrackB}
Track A tests whether the violations occur systematically, whereas Track B examines their mechanisms using eight hand-built diagnostics: three inverse pairs plus the broker and size-scaling constructions. Within each inverse pair the two constructions are structural contrasts designed to isolate the same confound from opposite directions. Opinion assignments are fixed except in T8, which runs Friedkin-Johnsen dynamics to equilibrium so that its two size-scaled blocks differ by size alone rather than by sampling. Table~\ref{tab:topologies} (appendix) states each construction, the confound it isolates, the behaviour theory predicts, and its role in the audit. The eight graphs are shown grouped by pair in the appendix (Figure~\ref{fig:topologies}).

\paragraph{Diagnostic classification.}
We deliberately avoid a binary polarized/not-polarized ground-truth label: polarization is a contested construct, and asserting ground truth would presuppose the questions this paper raises. Instead we apply two statistics designed to determine that. The first is bimodality coefficient (BC) of \citet{Freeman2013} which asks whether a distribution has two peaks or one, with a threshold of $0.555$ (the value a flat distribution returns). The second is the inter-cluster distance which asks how far apart those two humps sit ($2\mu$ under the bimodal opinion assignments, the equilibrium standard deviation under Friedkin-Johnsen). We require both, because the BC on its own can be fooled by a single lopsided hump. A subgraph is \emph{polarization-consistent} only when it clears both thresholds.
 
\paragraph{Statistical protocol.}\label{sec:statisticalevaluationprotocol}
Each check has a prespecified flagging rule. All $p$-values are Benjamini-Yekutieli corrected across the full family of tests; significance is treated as necessary but not sufficient, and effect sizes are reported throughout.
\begin{description}
\item[Measurement invariance] A two-sample Kolmogorov-Smirnov comparison of scores on polarization-consistent majority versus minority subgraphs (Track A). \emph{Flagged when} the difference is significant with Cohen's $|d|>0.3$.
\item[Structural exogeneity] \emph{Flagged when} a predictor's semi-partial $R^2$ exceeds $0.02$.
\item[Resolution sensitivity] The detection rate: the fraction of polarization-consistent minority subgraphs scoring above the T6 noise floor plus half a standard deviation, as a function of size ratio $r$. \emph{Flagged when} it falls below $0.8$ for any $r<0.5$.
\end{description}

\subsection{Real-World Experiments} \label{sec:realWorldExperiments}
 
\subsubsection{Dataset Selection and Scope}
The real-world phase is an external plausibility check: only the synthetic tracks supply the controlled conditions that separate measurement failure from genuine polarization differences. Two requirements govern selection: a network's polarization status must be established by prior analysis rather than by the measures under evaluation (else selection is circular), and it must carry scalar opinions or documented opinion proxies. We prefer widely cited benchmarks whose polarization status is a matter of disciplinary consensus.
 
\subsubsection{Datasets}
\textbf{US Congress roll-call voting} (Voteview \citep{Lewis2023}; DIME campaign-finance scores \citep{Bonica2014}). Legislators are nodes, edges weighted by roll-call voting-agreement rate. Each legislator's opinion score is their DIME/CFscore. The documented secular rise in congressional ideological distance gives a natural low-to-high ordering of snapshots, one per Congress across the DIME period; because the two parties sit near parity, Congress serves as the study's \emph{longitudinal} instrument rather than a source of size variation.
 
\textbf{European Parliament roll-call voting} (VoteWatch Europe \citep{VoteWatch2022} and HowTheyVote.eu \citep{HowTheyVote}; Chapel Hill Expert Survey \citep{CHES2024}). MEPs are nodes, edges constructed identically; each MEP carries the continuous left-right score of their national party from the corresponding CHES wave, expert-elicited and independent of both the votes and every measure under evaluation. Unlike near-parity, two-party Congress, the multiparty structure yields many minorities per snapshot spanning roughly three to twenty-five percent of seats, making the European Parliament the study's \emph{cross-sectional} instrument for whether minority scores co-vary with majority size.
 
We excluded two datasets one might expect: the Political Blogosphere \citep{Adamic2005}, whose binary labels and single snapshot cannot support the exogeneity regression, and the Twitter Controversy datasets \citep{Garimella2017}, whose community labels derive from a score in the Network-Weighted Distance's own family, making them circular for assessing NWD.
 
\subsubsection{Fairness Evaluation on Real-World Data}
For each dataset we score the full network, the majority, and every minority community. \textbf{Measurement invariance} uses a contribution ratio, the minority score relative to the majority score, against the ratio predicted by the minority's size and polarization intensity; a ratio systematically below expectation is a violation. \textbf{Structural exogeneity} tests whether a minority's score co-varies with majority size and density while its own parameters stay stable, across Congress snapshots (partialling out the longitudinal rise) and cross-sectionally across the European Parliament's coexisting groups; a significant partial correlation is a violation. \textbf{Resolution sensitivity} is descriptive: whether each minority score clears the synthetic T6 noise floor. For Congress we also test, via Kendall's $\tau$, whether each measure tracks the longitudinal rise consistently across majority and minority subgraphs.
 
\subsubsection{Stratification Exogeneity} \label{sec:stratificationExogeneity}
The factorial varies the majority's size, density, and connectivity, but not a different structural property: how strongly a network sorts into ordered status strata. A measure could be exogenous to majority size yet respond to a status hierarchy unrelated to the opinions being measured. The political panels cannot isolate this. Their only strong structural axis is the opinion axis itself, so we turn to a third network, the NetSense study \citep{NetSense}: roughly two hundred undergraduates, edges their call and text interactions aggregated per semester via the CogsNet model \citep{CogsNet}. Six survey-elicited worldview items form the opinion signal, independent of the graph; parents' income, on a fourteen-point scale, is the exogenous status attribute, fixed before the network forms.
 
We manipulate the stratification assortativity $\mathrm{StA}$ of \cite{Jalali2023}, a scalar in $[-1,1]$ for how strongly a network segregates into ordered tiers on an attribute, in three steps. A \emph{decoupling screen} keeps only worldview items independent of income (Spearman $|\rho|<0.2$), so a measure's response to income-stratification cannot be a legitimate opinion reading. Then, holding every opinion fixed, \emph{degree-preserving biased edge swaps} sweep $\mathrm{StA}$ on income from near zero to its maximum, leaving the degree sequence and all opinions unchanged. Finally each score is regressed on $\mathrm{StA}$: a stratification-exogenous measure stays flat, and one whose score climbs (semi-partial $R^2 > 0.02$) reports polarization the opinions do not contain. Structural modularity $Q_\text{struct}$ is the positive control and Esteban-Ray the negative one.
 
\subsubsection{Reproducibility}
Code and data are archived in a CodeOcean capsule (\url{https://codeocean.com/capsule/1963724/tree/v1}) and a GitHub repository (\url{https://github.com/StanislawStepien/fairness_of_polarization_measures}), with a fixed random seed making every reported value reproducible.

\section{Results}\label{sec:results}

No measure passes every applicable criterion across all tracks (Table~\ref{tab:audit_overview}); each is flagged on at least one. Contextual dependence appears through two dissociable channels: an opinion-formation channel that affects every opinion-reading measure and a stratification channel to which the opinion-free measure succumbs. Opinion dependence has opposite effects across the two channels: it transmits contextual dependence when context changes opinions, but can protect against opinion-orthogonal structural variation when opinions are fixed.

\subsection{Track A: Parametric Synthetic Evaluation}
\label{sec:results_trackA}

\begin{figure*}[t]
\centering
\includegraphics[width=\textwidth]{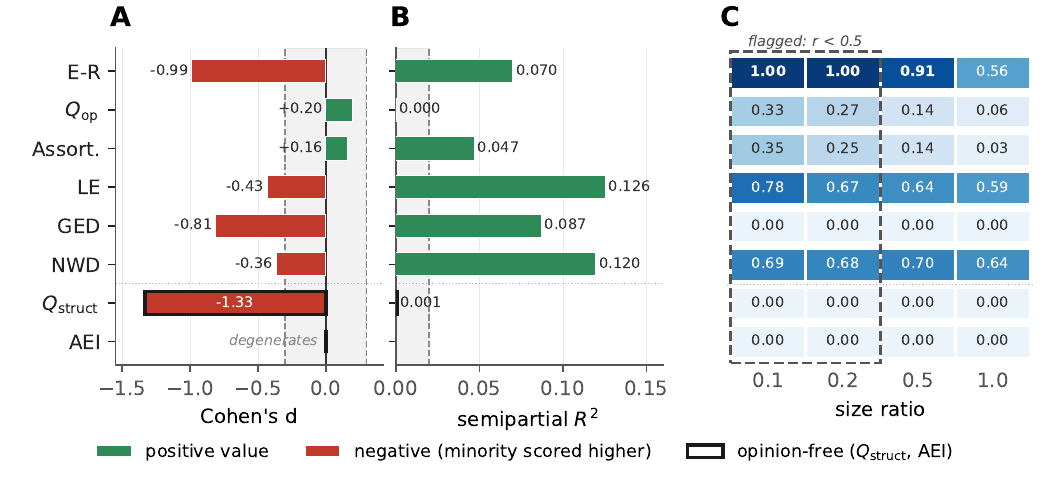}
\caption{The two fairness criteria and resolution-sensitivity prerequisite on Track A, per measure, ordered so opinion-reading measures sit above the two opinion-free ones ($Q_\text{struct}$, AEI). \textbf{(A)} Measurement invariance: Cohen's $d$ (majority $-$ minority); red bars are negative (the minority scored higher), and bars past $\pm0.3$ (dashed) are Benjamini-Yekutieli-significant violations. \textbf{(B)} Structural exogeneity: semi-partial $R^2$ of the minority score on majority parameters under Friedkin-Johnsen opinions, violations past $0.02$ (dashed); under fixed bimodal opinions, the control condition, all measures are exogenous ($R^2\le0.01$). \textbf{(C)} Resolution sensitivity: detection rate by size ratio $r$, one cell per measure, darker is higher and bold marks detection $\ge0.8$; the dashed box is the flagged region ($r<0.5$). AEI is degenerate on the single-community minority subgraphs, so its MI and SE entries are not substantively evaluable.}
\label{fig:criteria}
\end{figure*}

\subsubsection{Measurement Invariance}
\label{sec:results_MI}
Figure~\ref{fig:criteria}A reports Cohen's $d$ (majority $-$ minority) on structurally equivalent SBM subgraphs; a measure passes MI when $|d| < 0.3$ and the KS test is non-significant after Benjamini-Yekutieli correction. Five of the eight evaluated measures violate MI, and every violation runs the same way: the minority is scored higher. Structural modularity shows the largest effect ($d=-1.33$), followed by Esteban-Ray ($d=-0.99$), GED ($d=-0.81$), Laplacian Energy ($d=-0.43$), and NWD ($d=-0.36$). The modularity effect reflects the resolution limit; E-R's arises because the bimodal construction gives the two groups different opinion distributions. Opinion modularity ($d = +0.20$) and Assortativity ($d = +0.16$) stay within tolerance. AEI is degenerate on the single-community minority subgraphs and is therefore not counted as a substantive pass.

\subsubsection{Structural Exogeneity}
\label{sec:results_SE}
We regress each measure's minority-subgraph score on the majority parameters ($\log n_\text{maj}$, $p_\text{in}^\text{maj}$, $p_\text{out}$) under two opinion schemes and report the semi-partial $R^2_\text{maj}$ (Table~\ref{tab:se_betas}, Figure~\ref{fig:criteria}B). Under fixed (bimodal) opinions, the control condition in which majority structure is absent from the minority's scoring input, every measure is exogenous as expected: majority parameters explain at most about one percent of minority-score variance, indistinguishable from sampling noise. Under Friedkin-Johnsen dynamics, where the majority reaches the minority through the opinions it induces at equilibrium, five measures acquire a moderate but statistically robust dependence: Laplacian Energy ($R^2 = 0.126$) and NWD ($0.120$) most, then GED ($0.087$), Esteban-Ray ($0.070$) and Assortativity ($0.047$). The dependence runs through cross-group connectivity $p_\text{out}$ for E-R, Assortativity and LE, and through majority size for GED and NWD; both modularity operationalizations stay exogenous ($R^2 \approx 0.00$), and AEI is degenerate here. Because the contamination is located in opinion formation rather than in any measure's arithmetic, switching to a different opinion-reading measure does not remove it (the Discussion); the complementary T8 size-scaling diagnostic produces the same qualitative dependence under FJ-generated opinions.

\begin{table}[ht]
\centering
\caption{Structural Exogeneity, opinion-dynamics (Friedkin-Johnsen) channel: semi-partial $R^2_\text{maj}$ of each measure's minority-subgraph score on the majority parameters, the dominant majority predictor, and the violation verdict (Benjamini-Yekutieli-corrected). Under fixed (bimodal) opinions, the control condition, every measure is exogenous ($R^2 \le 0.01$, not shown). The Adaptive E-I Index is degenerate on single-community minority subgraphs.}
\label{tab:se_betas}
\small
\resizebox{\columnwidth}{!}{%
\begin{tabular}{|l|c|c|c|}
\hline
\rowcolor{nslblue!15}
Measure & $R^2_\text{maj}$ (FJ) & Dominant predictor & SE (FJ) \\
\hline
$Q_\text{opinion}$ & $0.00$ & - & \checkmark \\
\hline
$Q_\text{struct}$  & $0.00$ & - & \checkmark \\
\hline
E-R   & $0.070$ & $p_\text{out}$ & $\times$ \\
\hline
$A$   & $0.047$ & $p_\text{out}$ & $\times$ \\
\hline
LE    & $0.126$ & $p_\text{out}$ & $\times$ \\
\hline
GED   & $0.087$ & $\log n_\text{maj}$ & $\times$ \\
\hline
NWD   & $0.120$ & $\log n_\text{maj}$ & $\times$ \\
\hline
AEI   & \multicolumn{3}{c|}{degenerate on single-community subgraphs} \\
\hline
\end{tabular}%
}
\end{table}

\subsubsection{Resolution Sensitivity}
\label{sec:results_RS}
Figure~\ref{fig:criteria}C shows detection rate against size ratio $r = n_\text{min}/n_\text{maj}$; a measure is flagged where its rate falls below $0.8$ among polarization-consistent minority subgraphs. Fourteen measure $\times$ size-ratio pairs are flagged, all at $r \in \{0.1, 0.2\}$. Esteban-Ray alone passes throughout, a topology-blind distributional measure being immune to the resolution limit. The rest all fail at the two smallest ratios: structural modularity detects the minority in none of the small-ratio trials ($0.00$) and opinion modularity in about a third ($0.33$ at $r = 0.1$), AEI and GED never clear the noise floor ($0.00$), and Assortativity ($0.35/0.25$), Laplacian Energy ($0.78/0.67$) and NWD ($0.69/0.68$) detect it in most trials but still below the bar. Every structural measure dropping below reliable detection at $r \le 0.2$ points to a general minimum-group-size requirement for trustworthy polarization measurement.

\subsection{Track B: Diagnostic Topology Audit}
\label{sec:results_trackB}
\begin{figure}[ht]
    \centering
    \includegraphics[width=\linewidth]{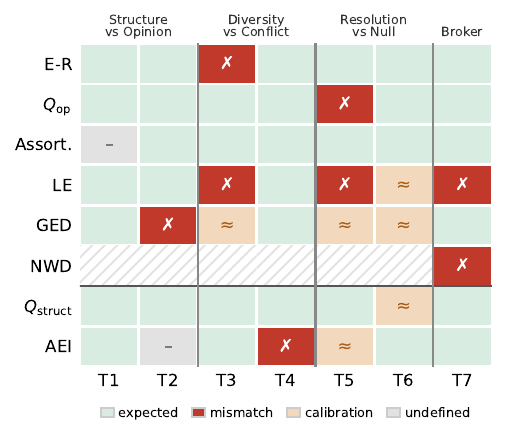}
    \caption{Track B diagnostic audit: which measures fail which diagnostic constructions. Each cell records whether a measure follows the theoretically expected direction on the corresponding topology; red marks a mismatch, amber a calibration or baseline artifact, gray a case undefined for that measure, and hatching a case not evaluated. The columns are grouped as T1/T2 \emph{Structure vs.\ Opinion}, T3/T4 \emph{Diversity vs.\ Conflict}, T5/T6 \emph{Resolution vs.\ Null}, and T7 \emph{Broker Dilution}. NWD is evaluated only on T7, the purpose-built broker-dilution diagnostic for its walk-based mechanism; the remaining diagnostics were not assigned prespecified directional predictions for NWD. Full constructions, scores, and expected directions are in the appendix (Table~\ref{tab:trackb_diagnostic}).}

    \label{fig:trackb_summary}
\end{figure}

Table~\ref{tab:trackb_diagnostic} (appendix) gives the direction check for every (topology, measure) pair; Figure~\ref{fig:trackb_summary} shows the pattern across the three inverse pairs and the broker diagnostic.

T1/T2 separate structural from opinion-based polarization; T3/T4 expose the distinction between diversity and conflict; and T5/T6 probe subgroup resolution and the null floor. T7 isolates broker dilution, under which NWD under-reports experienced opinion distance while GED remains robust. The size-only T8 construction provides a complementary robustness check: six measures show significant majority--minority differences under 10:1 scaling, whereas $Q_\text{opinion}$ does not (Figure~\ref{fig:t8_mean_diff}). Full topology constructions and per-measure scores are given in the appendix.

\begin{figure}[ht]
    \centering
    \includegraphics[width=0.85\linewidth]{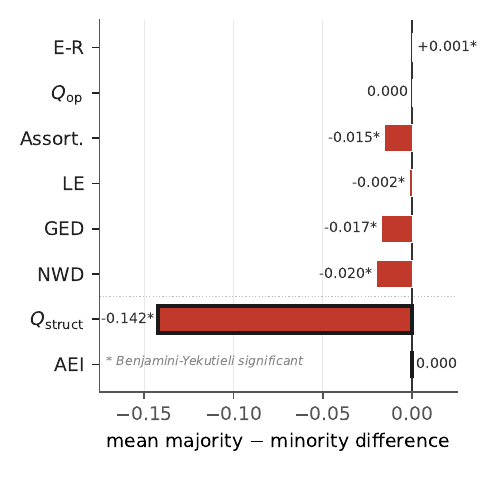}
    \caption{T8 (Asymmetric Echo Chambers): mean majority $-$ minority score difference across 100 replicate pairs. Asterisks mark Benjamini-Yekutieli-significant violations.}
    \label{fig:t8_mean_diff}
\end{figure}

\subsection{Real-World Plausibility Checks}
\label{sec:results_realworld}

\subsubsection{US Congress / CFscore}
On the thirteen-Congress panel (106th through 118th House, majority alternating between the two parties), the exogeneity test recovers the synthetic size channel. After partialling out the secular rise in polarization, three measures show significant negative partial correlations between minority score and majority size: Laplacian Energy ($r=-0.66$, $p=0.014$), GED ($r=-0.65$, $p=0.016$), and NWD ($r=-0.62$, $p=0.025$). These are the three strongest synthetic leaks through the opinion-formation channel, linking the controlled and observational analyses. Structural modularity ($r=0.08$), Esteban-Ray ($r=0.27$), and Assortativity ($r=-0.02$) show no size dependence. Opinion modularity is the lone discrepancy ($r=0.57$, $p=0.042$), plausibly because in real data majority size is entangled with partisan era and minority composition, unlike the controlled sweep. Under MI, Esteban-Ray most consistently under-credits the minority (nine of thirteen Congresses); the longitudinal tracking test is inconclusive on this short panel ($|\tau|\le0.32$, $p>0.13$). Resolution sensitivity is not assessed because the real-world panels lack the synthetic noise-floor reference.

\subsubsection{European Parliament / CHES}
The European Parliament panel provides weaker, inconclusive evidence. The exogeneity test returns a uniform null (all $|r|\le0.18$, $p\ge0.35$), but only four distinct majority sizes are observed and the same centre-right majority family persists across terms. A secondary within-term analysis is also mostly null; its one nominally significant result (Esteban-Ray, $r=0.47$, $p=0.02$) does not survive correction. Therefore, the European Parliament provides results consistent with the size channel, though not strong evidence for it.

\subsubsection{Stratification exogeneity on the NetSense communication network}\label{sec:results_sta}
Across the five per-semester NetSense networks (S2 through S6, each with roughly one hundred and ninety participants carrying income data), parents' income is decoupled from every worldview item (income-opinion rank correlation never exceeding $0.19$ across all thirty semester-item pairs), so income provides an approximately opinion-orthogonal status attribute for this analysis. $Q_\text{struct}$ serves as a positive control: its score rises sharply with injected income stratification ($R^2$ between $0.65$ and $0.80$). The substantive result is that the opinion-reading measures remain largely invariant to the same rewiring: Esteban-Ray has exactly zero variance across the sweep, and GED and the spectral and relational measures sit at or near the $0.02$ floor, rising above it only sporadically for particular items (GED reaches about $0.12$ in one semester, on a single welfare-spending item). Here, reading opinions protects against the injected structural signal: the opinion-free measure is the one that mistakes stratification for polarization. The stratification is modest ($\mathrm{StA}$ near $0.17$ across all five semesters), so the probe characterises each measure's response function rather than asserting strong stratification in these particular networks.

\subsection{Cross-Measure Audit Summary}
\label{sec:results_summary}
Table~\ref{tab:audit_overview} consolidates the per-criterion verdicts across all tracks. No measure passes every applicable criterion, and resolution sensitivity limits the interpretable operating range of several measures; a resolution-sensitivity flag means a measure cannot be interpreted at that scale rather than that it is unfair in itself.

\begin{table}[t]
\centering
\caption{Cross-measure audit overview, one column per check. MI (Measurement Invariance) and SE$_\text{FJ}$ (Structural Exogeneity via the Friedkin-Johnsen channel) are the fairness checks; RS (Resolution Sensitivity) is a validity prerequisite, and T8 is a size-scaling diagnostic; TB is the Track B mismatch count. \checkmark\ = pass, $\times$ = violation, - = not meaningfully evaluable. $^\dagger$ The Adaptive E-I Index is degenerate on the single-community minority subgraphs, so its MI and SE$_\text{FJ}$ entries are definitional rather than substantive.}
\label{tab:audit_overview}
\small
\setlength{\tabcolsep}{4pt}
\begin{tabular}{|l|c|c|c|c|c|}
\hline
\rowcolor{nslblue!15}
Measure & MI & SE$_\text{FJ}$ & T8 size & RS & TB \\ \hline
$Q_\text{opinion}$ & \checkmark & \checkmark & \checkmark & $\times$ & 1 \\ \hline
$Q_\text{struct}$  & $\times$ & \checkmark & $\times$ & $\times$ & 0 \\ \hline
E-R   & $\times$ & $\times$ & $\times$ & \checkmark & 1 \\ \hline
$A$   & \checkmark & $\times$ & $\times$ & $\times$ & 0 \\ \hline
LE    & $\times$ & $\times$ & $\times$ & $\times$ & 3 \\ \hline
GED   & $\times$ & $\times$ & $\times$ & $\times$ & 1 \\ \hline
AEI   & -$^\dagger$ & -$^\dagger$ & \checkmark & $\times$ & 1 \\ \hline
NWD   & $\times$ & $\times$ & $\times$ & $\times$ & 1 \\ \hline
\end{tabular}
\end{table}

\section{Discussion}
\label{sec:discussion}

Our results should not be read as an argument against the use of the tested measures - each remains a reasonable tool built to capture aspects of polarization. What the audit adds is a caution about interpretation. When such a measure is applied to a minority community embedded in a larger network whose opinions formed through social influence, the score it assigns the minority reflects, in part, the surrounding majority, not the minority alone: a group is described by a number shaped by a structural context it did not choose.

\subsection{Structural exogeneity fails through two dissociable channels}\label{sec:disc_se}

Because each minority is scored on its own subgraph, the majority can reach that score only through the opinions the minority holds. The bimodal condition is a control that confirms this: with opinions drawn independently of the network, majority structure is absent from the minority's input, and the near-zero dependence we observe (at most about one percent of variance, indistinguishable from noise) shows the pipeline adds no spurious signal. Under Friedkin-Johnsen equilibrium opinions \citep{FriedkinJohnsen1990}, where the majority reaches the minority by shaping the opinions it holds at equilibrium, the distributional, relational, and spectral measures acquire a moderate dependence, Laplacian Energy and the Network-Weighted Distance most and both modularity operationalizations still exogenous, though even the largest leaves no more than roughly one eighth of a measure's variance to majority structure (Table~\ref{tab:se_betas}). Since the minority's structure and internal ties are held fixed, this dependence can enter only through opinion formation, not the arithmetic of any measure, so within this channel it cannot be removed by switching measures: every opinion-reading measure inherits the same contaminated opinions.

The second channel is different because the dependence arises in the measure itself and can therefore be avoided by choosing a different measure. Holding every opinion fixed but rewiring the NetSense communication network to raise its stratification assortativity \citep{Jalali2023} on parents' income, approximately uncorrelated with every worldview item ($|\rho| < 0.2$), a fair measure should not move. Two facts make this more than the observation that structural measures track structure. Our factorial rates $Q_\text{struct}$ exogenous ($R^2 \approx 0.0005$), yet under stratification its score follows the induced structure at $R^2$ between $0.65$ and $0.80$: the factorial is not wrong, only blind to a channel it never varies. And the failure tracks exactly whether a measure reads opinions: the opinion-only Esteban-Ray index stays invariant and every opinion-reading measure largely resists, while the opinion-free modularity succumbs wholesale.
The two experiments point in different directions. If the surrounding network changes opinions, every opinion-based measure inherits that change. If the network changes while opinions remain fixed, however, using the opinions can insulate a measure from the structural perturbation. Thus contextual robustness is not predicted by dimensional breadth, but by how a measure uses opinions.

The stratification interpretation assumes polarization is opinion-based rather than segregation-based; under the latter definition, $Q_\text{struct}$'s response is expected.

\subsection{Measurement invariance and the detectability prerequisite}
\label{sec:disc_mi_rs}

Measurement invariance asks a related question: do a majority and minority in structurally equivalent positions receive equivalent scores? Five of the eight evaluated measures show a significant gap (Cohen's $|d| > 0.3$ after correction), and in every case the minority is scored \emph{higher}, read as more polarized than an identically configured majority. Only $Q_\text{opinion}$ and Assortativity stay within tolerance. The T8 size-scaling diagnostic, which compares proportionally identical blocks at different absolute sizes, registers a significant majority-minority gap for six of the eight measures, largest by far for structural modularity, while $Q_\text{opinion}$ does not. We report it here rather than under structural exogeneity because it compares two co-existing groups rather than varying a majority against a fixed minority.

Because the criteria are not independent, we treat a resolution-sensitivity failure as disqualifying: a measure that cannot detect a minority returns a near-constant score - its invariance and exogeneity clearances are vacuous. The Adaptive E-I Index is the clearest case, its near-zero effect size a product of degeneracy on the single-community minority subgraphs. Both modularity operationalizations fail the resolution check at the two smallest size ratios and clear it only at $0.5$ and above \citep{Fortunato2007}, so neither can be read as fair at those scales whatever else it clears.

\subsection{Dimension coverage in light of the results}
\label{sec:disc_dimensions}

Dimensional breadth, the criterion behind our measure selection, does not predict fairness. If it did, the widest measure would be the least fair. Instead the two largest invariance violations are single-dimension measures (Esteban-Ray and structural modularity, both exceeding the all-dimension GED), and, most sharply, the two modularities share a single dimension yet diverge by an order of magnitude on the stratification probe ($Q_\text{struct}$ at $R^2 = 0.71$, $Q_\text{opinion}$ near $0.06$). Coverage describes what a measure captures, but only whether it reads opinions predicts how it behaves.

\subsection{Practical guidance}
\label{sec:disc_guidance}

No measure needs to be abandoned. The concern is narrower: minority polarization scores need care whenever the minority sits in a larger network with socially formed opinions. We therefore recommend reporting the surrounding majority's size and density alongside any minority's polarization value. Comparisons made \emph{within} a single network are preferable to absolute scores across settings, a minority's absolute score is better read as context-relative than intrinsic, and a measure should be confirmed to resolve the minority at its size ratio before its score is interpreted. Two cautions are measure-specific: Laplacian Energy and the Network-Weighted Distance leak most through the opinion-formation channel, while Assortativity is the most size-sensitive. On an embedded minority, all three should therefore be read against a fixed majority context rather than compared across settings.

\subsection{Limitations and future work}
\label{sec:disc_limitations}

We treat sensitivity to majority-induced changes in minority opinions as a fairness concern, whereas a reader might regard such influence as a measure working as intended. The practical advice does not depend on resolving that disagreement: it concerns comparability rather than whether the resulting opinion changes are desirable. Even under the alternative interpretation, the dependence itself remains.

Three boundaries remain. Our design scores minorities in isolation, so it does not test direct effects of boundary edges on a fixed minority; a boundary-inclusive design would address this channel. The opinion-formation result is specific to Friedkin--Johnsen dynamics and may attenuate under bounded-confidence dynamics \citep{Deffuant2000, Hegselmann2002}. Finally, the effect-size and detection-rate thresholds are pragmatic rather than derived; the underlying effect sizes are reported so readers can reassess them.

\section*{Funding}
This work was supported by the National Science Centre, Poland, under Grant 2021/41/B/HS6/02798, and by the Polish Ministry of Education and Science within the program ``International Projects Co-Funded''.

{\small
\bibliographystyle{plainnat}
\bibliography{bibliography}
}

\appendix

\subsection*{Appendix: Synthetic experiment parameters}

\paragraph{Structural exogeneity factorial.}
Minority held fixed at $n_\text{min} = 100$, $p_\text{in}^\text{min} = 0.3$. Majority swept independently over $n_\text{maj} \in \{200, 500, 1000, 2000\}$, $p_\text{in}^\text{maj} \in \{0.05, 0.15, 0.30, 0.50\}$, and $p_\text{out} \in \{0.001, 0.005, 0.02\}$. Each cell is run under both the bimodal and the Friedkin–Johnsen equilibrium opinion schemes, with 50 replicates per cell.

\paragraph{Track A network families.}
Erdős–Rényi $G(n,p)$: $n = 500$, $p$ calibrated to mean degree $\langle k \rangle \in \{5, 10, 20\}$. Barabási–Albert: attachment parameter $m \in \{2, 5, 10\}$. Stochastic block model: two blocks, $p_\text{in} = 0.15$, $p_\text{out} = 0.015$, size ratio $r = n_\text{min}/n_\text{maj} \in \{0.1, 0.2, 0.5, 1.0\}$.

\paragraph{Opinion assignment schemes.}
Bimodal: two-Gaussian mixture at $\pm\mu$ with $\sigma = 0.1$, for $\mu \in \{0.4, 0.6, 0.8\}$. FJ-equilibrium: $\alpha \in \{0.3, 0.6, 0.9\}$, stubborn vector $\mathbf{s} \sim \mathcal{N}(\pm\mu_s, \sigma_s)$ bi-modally initialized by block (randomly in Erdős–Rényi and BA). Each network-assignment combination replicated 50 times with independent seeds.

\section{Track B diagnostic detail}\label{app:trackb}

\begin{figure*}[t]
    \centering
    \includegraphics[width=\textwidth,height=0.75\textheight,keepaspectratio]{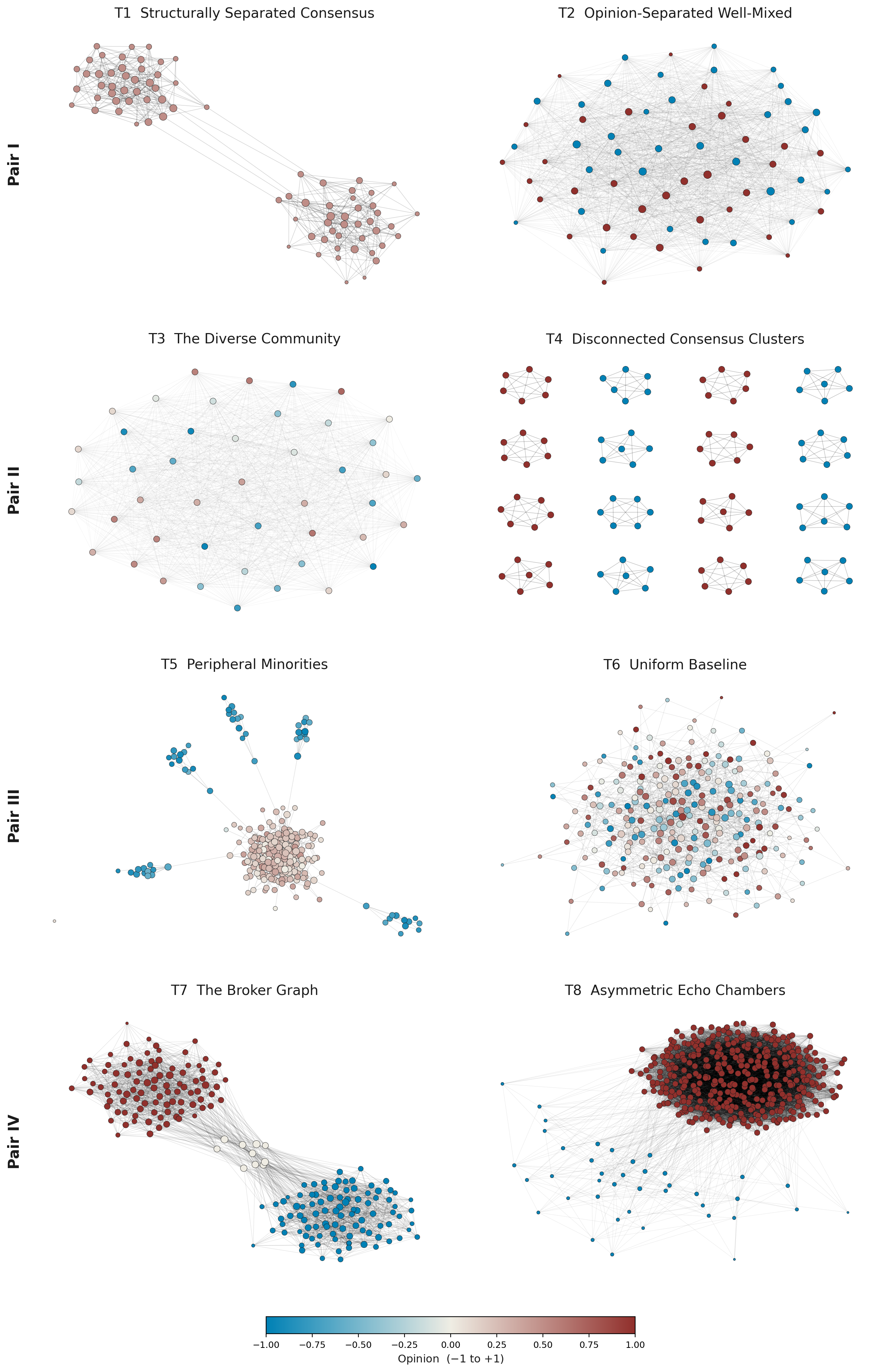}
    \caption{\textbf{Diagnostic topologies (Track B).} Rows are the three inverse pairs and two standalone diagnostics; within each inverse pair the two topologies probe a measure from opposite directions. Node color encodes opinion on a fixed scale (blue $-1$, cream $0$, dark red $+1$); layout and node size are illustrative. Construction and expected behaviour are given in Table~\ref{tab:topologies}.}
    \label{fig:topologies}
\end{figure*}

Track B uses three inverse pairs and two standalone diagnostics. Figure~\ref{fig:topologies} shows the constructions, Table~\ref{tab:topologies} gives their design logic and expected behaviour, and Table~\ref{tab:trackb_diagnostic} reports the per-measure results.

\begin{table*}[t]
\centering
\caption{The eight diagnostic topologies (Track B), grouped into three inverse pairs and two standalone diagnostics. Each pair probes one measure from opposite directions; the final column records the role of each construction in the audit. Full generation parameters are given in the appendix.}
\label{tab:topologies}
\begin{tabular}{|L{2.4cm}|L{3.2cm}|L{2.1cm}|L{4.8cm}|L{1.7cm}|}

\hline
\rowcolor{nslblue!15}
Topology & Construction & Isolates & Predicted behaviour & Criterion \\ \hline
\rowcolor{nslblue!15}
\multicolumn{5}{|l|}{\textbf{Pair I - Structure without opinion vs.\ opinion without structure} (probes Q, GED)} \\ \hline
T1 Separated Consensus & Two dense communities, all $x=+0.5$ & structure, no opinion & Q reports high (structural false positive); E-R and LE correctly $\approx 0$ & SE \\ \hline
T2 Opinion-Separated Mixed & Dense E-R ($p=0.7$), bimodal $\pm 1$, edges opinion-blind & opinion, no structure & Q $\approx 0$ (miss); E-R max; LE high; GED should be nonzero & coverage \\ \hline
\rowcolor{nslblue!15}
\multicolumn{5}{|l|}{\textbf{Pair II - Diversity vs.\ conflict} (probes LE, GED)} \\ \hline
T3 Diverse Community & One clique, $x \sim \mathcal{U}(-1,1)$ & diversity, no separation & LE high (false positive from density $\times$ range); GED suppresses via its mesolevel term & coverage \\ \hline
T4 Disconnected Clusters & 20 disconnected cliques, alternating $\pm 1$ & distributional separation, no cross-edges & LE exactly $0$ (miss); E-R max & coverage \\ \hline
\rowcolor{nslblue!15}
\multicolumn{5}{|l|}{\textbf{Pair III - Peripheral minorities and the noise floor} (probes Q, AEI, A, E-R)} \\ \hline
T5 Peripheral Minorities & $0.8N$ majority + five $n{=}20$ minority clusters, one bridge edge each & small peripheral minorities & Q misses them (resolution limit); A majority-dominated; E-R cannot tell five isolated groups from one bloc & RS \\ \hline
T6 Uniform Baseline & E-R $n{=}500$, $\langle k \rangle{=}10$, $x \sim \mathcal{U}(-1,1)$ & neither structure nor clustering & defines the empirical noise floor against which all scores are read & baseline \\ \hline
\rowcolor{nslblue!15}
\multicolumn{5}{|l|}{\textbf{Broker dilution diagnostic} (probes NWD, GED, LE)} \\ \hline
T7 Broker Graph & Two dense $\pm 1$ communities joined \emph{only} through moderate ($x{=}0$) high-degree brokers & experienced vs.\ structural & NWD under-reports (brokers dilute the aggregate), so a community reached mainly through brokers is under-scored; whether GED is diluted likewise is open & SE \\ \hline
\rowcolor{nslblue!15}
\multicolumn{5}{|l|}{\textbf{Size-scaling diagnostic} (probes MI)} \\ \hline
T8 Asymmetric Echo Chambers & Two proportionally identical SBM blocks, $n_\text{maj}{=}10\,n_\text{min}$, FJ-equilibrium opinions; $\times 100$ replicates & size-scaled blocks with FJ-generated opinions & a majority-minority gap indicates size-dependent contextual sensitivity, mediated by the opinions induced by FJ dynamics. & MI / size diagnostic \\ \hline
\end{tabular}%

\end{table*}

\begin{table*}[t]
\centering
\caption{Track B diagnostic results, reproduced from the pipeline output. Cell shading: green = score matched its theoretically expected direction; red = mismatch; amber = calibration finding (MOD/NONZERO) or undefined (N/A). Scores are on the pipeline's normalized scale.}
\label{tab:trackb_diagnostic}
\scriptsize
\setlength{\tabcolsep}{3pt}
\renewcommand{\arraystretch}{0.95}
\resizebox{\textwidth}{!}{%
\begin{tabular}{|l|l|r|l|l|l|}
\hline
\rowcolor{nslblue!15}
Topology & Measure & Score & Expected & Direction check & Role \\ \hline
T1 & $Q_\text{opinion}$ & $0.0000$ & ZERO & \cellcolor{nslgreen!14}ok &  \\ \hline
T1 & $Q_\text{struct}$ & $0.4954$ & HIGH & \cellcolor{nslgreen!14}ok & MI \\ \hline
T1 & E-R & $0.0000$ & ZERO & \cellcolor{nslgreen!14}ok &  \\ \hline
T1 & $A$ & - & N/A & \cellcolor{nslorange!16}N/A &  \\ \hline
T1 & LE & $0.0000$ & ZERO & \cellcolor{nslgreen!14}ok &  \\ \hline
T1 & GED & $0.0000$ & ZERO & \cellcolor{nslgreen!14}ok &  \\ \hline
T1 & AEI & $0.9959$ & HIGH & \cellcolor{nslgreen!14}ok &  \\ \hline
T2 & $Q_\text{opinion}$ & $-0.0054$ & ZERO & \cellcolor{nslgreen!14}ok &  \\ \hline
T2 & $Q_\text{struct}$ & $0.0423$ & ZERO & \cellcolor{nslgreen!14}ok & MI \\ \hline
T2 & E-R & $0.3134$ & HIGH & \cellcolor{nslgreen!14}ok &  \\ \hline
T2 & $A$ & $-0.0108$ & ZERO & \cellcolor{nslgreen!14}ok &  \\ \hline
T2 & LE & $1.0107$ & HIGH & \cellcolor{nslgreen!14}ok &  \\ \hline
T2 & GED & $1.4401$ & HIGH & \cellcolor{nslred!20}\textbf{MISMATCH} &  \\ \hline
T2 & AEI & - & N/A & \cellcolor{nslorange!16}N/A &  \\ \hline
T3 & $Q_\text{opinion}$ & $-0.0050$ & ZERO & \cellcolor{nslgreen!14}ok &  \\ \hline
T3 & $Q_\text{struct}$ & $0.0000$ & ZERO & \cellcolor{nslgreen!14}ok &  \\ \hline
T3 & E-R & $0.0070$ & LOW & \cellcolor{nslred!20}\textbf{MISMATCH (not low)} &  \\ \hline
T3 & $A$ & $-0.0101$ & ZERO & \cellcolor{nslgreen!14}ok &  \\ \hline
T3 & LE & $0.2995$ & HIGH & \cellcolor{nslred!20}\textbf{MISMATCH} & MI \\ \hline
T3 & GED & $0.2965$ & MOD & \cellcolor{nslorange!16}MOD (baseline artifact) &  \\ \hline
T3 & AEI & $0.0000$ & ZERO & \cellcolor{nslgreen!14}ok &  \\ \hline
T4 & $Q_\text{opinion}$ & $0.5000$ & HIGH & \cellcolor{nslgreen!14}ok &  \\ \hline
T4 & $Q_\text{struct}$ & $0.9500$ & HIGH & \cellcolor{nslgreen!14}ok &  \\ \hline
T4 & E-R & $0.3134$ & HIGH & \cellcolor{nslgreen!14}ok &  \\ \hline
T4 & $A$ & $1.0000$ & HIGH & \cellcolor{nslgreen!14}ok &  \\ \hline
T4 & LE & $0.0000$ & ZERO & \cellcolor{nslgreen!14}ok & MI \\ \hline
T4 & GED & $-0.0000$ & ZERO & \cellcolor{nslgreen!14}ok & MI \\ \hline
T4 & AEI & $0.0000$ & HIGH & \cellcolor{nslred!20}\textbf{MISMATCH} &  \\ \hline
\end{tabular}%
}
\end{table*}

\begin{table*}[t]\ContinuedFloat
\centering
\caption{Track B diagnostic results (continued).}
\scriptsize
\setlength{\tabcolsep}{3pt}
\renewcommand{\arraystretch}{0.95}
\resizebox{\textwidth}{!}{%
\begin{tabular}{|l|l|r|l|l|l|}
\hline
\rowcolor{nslblue!15}
Topology & Measure & Score & Expected & Direction check & Role \\ \hline
T5 & $Q_\text{opinion}$ & $0.0303$ & ZERO & \cellcolor{nslred!20}\textbf{MISMATCH} & RS \\ \hline
T5 & $Q_\text{struct}$ & $0.0837$ & ZERO & \cellcolor{nslgreen!14}ok & RS \\ \hline
T5 & E-R & $0.0241$ & HIGH & \cellcolor{nslgreen!14}ok &  \\ \hline
T5 & $A$ & $0.6238$ & HIGH & \cellcolor{nslgreen!14}ok &  \\ \hline
T5 & LE & $0.0093$ & HIGH & \cellcolor{nslred!20}\textbf{MISMATCH} &  \\ \hline
T5 & GED & $1444.8210$ & MOD & \cellcolor{nslorange!16}MOD (baseline artifact) & RS \\ \hline
T5 & AEI & $0.9997$ & MOD & \cellcolor{nslorange!16}MOD (baseline artifact) & RS \\ \hline
T6 & $Q_\text{opinion}$ & $-0.0066$ & ZERO & \cellcolor{nslgreen!14}ok &  \\ \hline
T6 & $Q_\text{struct}$ & $0.2765$ & NONZERO & \cellcolor{nslorange!16}NONZERO (baseline artifact) &  \\ \hline
T6 & E-R & $0.0057$ & ZERO & \cellcolor{nslgreen!14}ok &  \\ \hline
T6 & $A$ & $0.0084$ & ZERO & \cellcolor{nslgreen!14}ok &  \\ \hline
T6 & LE & $0.3320$ & NONZERO & \cellcolor{nslorange!16}NONZERO (baseline artifact) &  \\ \hline
T6 & GED & $20.2994$ & NONZERO & \cellcolor{nslorange!16}NONZERO (baseline artifact) &  \\ \hline
T6 & AEI & $0.0000$ & ZERO & \cellcolor{nslgreen!14}ok &  \\ \hline
T7 & $Q_\text{opinion}$ & $0.4322$ & HIGH & \cellcolor{nslgreen!14}ok &  \\ \hline
T7 & $Q_\text{struct}$ & $0.4379$ & HIGH & \cellcolor{nslgreen!14}ok &  \\ \hline
T7 & E-R & $0.2746$ & HIGH & \cellcolor{nslgreen!14}ok &  \\ \hline
T7 & $A$ & $0.9272$ & HIGH & \cellcolor{nslgreen!14}ok &  \\ \hline
T7 & LE & $0.0679$ & HIGH & \cellcolor{nslred!20}\textbf{MISMATCH} &  \\ \hline
T7 & GED & $131.7666$ & HIGH & \cellcolor{nslgreen!14}ok &  \\ \hline
T7 & AEI & $0.6938$ & HIGH & \cellcolor{nslgreen!14}ok &  \\ \hline
T7 & NWD & $0.2359$ & LOW & \cellcolor{nslred!20}\textbf{MISMATCH (not low)} & MI \\ \hline
\end{tabular}%
}
\end{table*}

\paragraph{Reproducibility detail.}
The complete audit runs from \texttt{code/run.py}: experimental tracks together with the figures. Every table and figure is written to \texttt{results/}. Individual stages run in isolation through a \texttt{RUN\_ONLY} selector, and the measure implementations, the fairness criteria, the diagnostic-topology constructions, and the stratification probe are each isolated in their own module. The capsule mounts the real-world datasets as read-only assets and runs the full pipeline unattended; the NetSense per-semester networks are rebuilt from the raw log by \texttt{code/stratification/build\_netsense.py}.
 
\paragraph{European Parliament source combination.}
Roll-call records are taken from the VoteWatch Europe archive \citep{VoteWatch2022} across terms, with the more recent HowTheyVote.eu records \citep{HowTheyVote} (ninth and tenth terms) used to cross-check the ninth. Terms six through nine are matched to Chapel Hill Expert Survey waves 2006, 2010, 2014, and 2019 respectively. Combining sources trades some cross-term homogeneity of vote coding for the multi-term variation in majority size that the exogeneity test requires and that no single source spanning only two terms can supply; because the CHES left-right scores are expert-elicited at the party level rather than read off the votes, the crosswalk is robust to the change of roll-call source.

\end{document}